\documentstyle[aps,multicol,epsfig,epsf,psfrag]{revtex}

\draft
\newcommand{\beq}{\begin{equation}}
\newcommand{\eeq}{\end{equation}}
\newcommand{\bdis}{\begin{displaymath}}
\newcommand{\edis}{\end{displaymath}}
\newcommand{\bea}{\begin{eqnarray}}
\newcommand{\eea}{\end{eqnarray}}
\newcommand{\barr}{\begin{array}}
\newcommand{\earr}{\end{array}}
 
\begin{document}

\title{Universality classes in creep rupture}

\author{Ferenc
  Kun$^{1}$\footnote{Electronic
address:feri@dtp.atomki.hu}, Yamir Moreno$^{2}$, Raul
Cruz Hidalgo$^{3}$, Hans. J. Herrmann$^{3}$} 

\address{$^1$Department of Theoretical Physics, University of Debrecen, \\ 
P.O.Box: 5, H-4010 Debrecen, Hungary \\
$^2$The Abdus Salam International Center for Theoretical Physics
(ICTP), Condensed Matter Group, P.O.Box: 586, I-34014 Trieste, Italy, \\
$^3$Institute for Computational Physics, University of
Stuttgart,    
Pfaffenwaldring 27, 70569 Stuttgart, Germany\\
}

\date{\today}
\maketitle
\begin{abstract} 
We study the creep response of solids to a constant external load in
the framework of a novel fiber bundle model introduced. 
Analytical and numerical calculations showed that increasing the
external load on a specimen a transition takes place from a
partially failed state of infinite lifetime to a state where global
failure occurs at a finite time. 
Two universality classes of creep rupture were identified depending on
the range of interaction of fibers: in the  
mean field limit the transition between the two states is continuous
characterized by power law divergences, while for local interactions
it becomes abrupt with no scaling. Varying the range of interaction a
sharp transition is revealed between the mean field and short range
regimes. The creeping 
system evolves into a macroscopic stationary state accompanied by the
emergence of a power law distribution of inter-event times of the
microscopic relaxation process, which indicates self organized
criticality in creep.  
\end{abstract}


\begin{multicols}{2}
\narrowtext

Time evolution of physical systems under a steady external driving is
abundant in nature. Examples can be found in seemingly diverse systems
such as the behavior of domain walls in magnets \cite{zap98},
earthquake dynamics \cite{cha,tur97} and creep rupture of solids
\cite{exper1,guarino1,guarino2,aless,exper2,hirata}. Recent intensive
experimental and theoretical studies revealed that, in spite of the
diversity of the 
governing physical dynamics, the behavior of these systems share
several universal features: local stresses arising due to the external
driving are resolved in the form of avalanches of microscopic
relaxation events. As a result, the system attains a stationary
macroscopic state characterized by a scale invariant microscopic
activity. In particular, large efforts have been devoted to the study
of material failure occurring under various loading conditions
\cite{exper1,guarino1,guarino2,aless,exper2,hirata} since
these studies also provide direct information on the dynamics of
earthquakes \cite{cha,tur97,exper2,hirata}.

Under high steady stresses, materials may undergo time dependent
deformation resulting in failure called creep rupture which limits
their lifetime, and hence, has a high impact on their applicability in
construction elements. Creep failure tests are usually performed under
uniaxial tensile loading when the specimen is subjected to a constant
load $\sigma_{\rm o}$ and the time evolution of the damage process is
followed by recording the strain $\varepsilon$ of the specimen and the
acoustic signals emitted by microscopic failure events. 
In spite of the large amount of experimental results accumulated a
comprehensive theoretical picture of creep rupture is still lacking.

In this Letter we study the creep rupture of materials by means of
a novel fiber bundle model. During the past years Fiber Bundle Models (FBMs)
\cite{hansen,sornette3,kun1,kun4,yamir1,yamir2,chakrab,moral} of materials
damage played a very important role not only in the study of
fracture but they turned out to be also one of the
most promising approaches to earthquake predictions \cite{tur97,newzar}. 
Our model realistically describes the
interaction of fibers covering all cases relevant to real materials,
furthermore, it combines their viscoelastic constitutive behavior and
breaking. 
Analytical and numerical calculations show that there exists a
critical load that determines 
the final state of the material. Strikingly, we find that in creep
rupture there are only two universality classes and that the
distribution of times between breaking events is reminiscent of the
Omori's law in earthquake dynamics \cite{omori}. The results suggest
that the same dynamics holds from the scale of earthquakes down to the
rheological response of solids. The relevance of our results to
experiments is discussed.

Our model consists of $N$ parallel fibers arranged on a square lattice
of side length $L$ having viscoelastic constitutive behavior
\cite{kun4}.  For simplicity, the pure viscoelastic behavior of fibers
is modeled by a Kelvin-Voigt element which consists of a spring and a
dashpot in parallel and results in the constitutive equation
$\sigma_{\rm o} = \beta \dot{\varepsilon} + E\varepsilon$, where
$\sigma_{\rm o}$ is the imposed load, $\beta$ denotes the damping
coefficient, and $E$ the Young modulus of fibers, respectively.  In
order to capture failure in the model a strain controlled breaking
criterion is imposed, {\it i.e.} a fiber fails during the time
evolution of the system when its strain exceeds a breaking threshold
$\varepsilon_i, \ \ i=1, \ldots , N$ drawn from a probability
distribution $P(\varepsilon)=\int_0^{\varepsilon}p(x)dx$.  When a
fiber fails its load is redistributed to the intact fibers according
to the interaction law of fibers. It was shown in Ref. \cite{kun4}
that in a viscoelastic bundle the fibers break one-by-one
which implies that such a system is very sensitive to the details of
load sharing, much more than a static fiber bundle
\cite{hansen,sornette3,kun1,kun4,yamir1,yamir2,chakrab} where a large
number of fibers can break at once in the form of bursts
\cite{hansen}.  In order to realistically model the stress transfer
between fibers, recently an adjustable stress transfer function was
introduced, which interpolates between the two limiting cases of
global and local load sharing \cite{sornette3,kun1} (GLS and LLS).
Motivated by results of fracture mechanics we assume that the
additional load $\sigma_{add}$ received by an intact fiber $i$ on a
square lattice after the failure of fiber $j$ depends on their
distance $r_{ij}$ and has the form
$\sigma_{add}=Zr_{ij}^{-\gamma}$. Here the normalization factor $Z$ is
simply $Z=\sum_{i\in I} r_{ij}^{-\gamma}$ and the sum runs over the
set $I$ of intact fibers\cite{yamir2}.  The exponent $\gamma$ is an
adjustable parameter of the model, which controls the effective range
of load redistribution. The limiting cases $\gamma=0$ and $\gamma
\rightarrow \infty$ recover the global and local load sharing widely
studied in the literature, furthermore, intermediate values of
$\gamma$ interpolate between them. A comprehensive study of the
quasistatic fracture of fiber bundles in terms of the adjustable load
sharing function has been presented in Ref.\ \cite{yamir2}.

\begin{figure}[b]
\begin{center}
\epsfig{bbllx=140, bblly=140,bburx=475,bbury=655,file=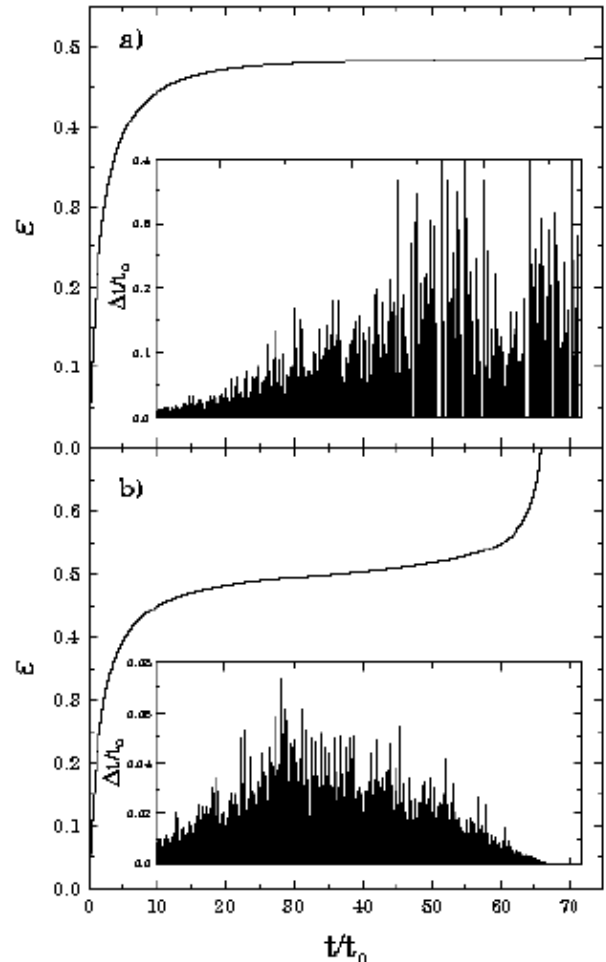,
  width=8.5cm}  
\caption{\small $\varepsilon(t)$ for two different values of the
  external load $a)$ $\sigma_o <\sigma_c$, $b)$ $\sigma_o
  >\sigma_c$. The insets present the inter-event times $\Delta t$ at
  the time of their occurrence $t$.} 
\label{fig:eps_t}
\end{center}
\end{figure}

In the mean field limit, {\it i.e.} global load
sharing obtained for $\gamma = 0$, many of the quantities
describing the behavior of the system can be obtained analytically. In
this case the time  
evolution of the system under a steady external load $\sigma_o$ is
described by the differential equation  
\begin{eqnarray}
  \label{eq:eom}
  \frac{\sigma_{\rm o}}{1-P(\varepsilon)} = \beta \dot{\varepsilon}
  +E\varepsilon, 
\end{eqnarray}
where the viscoelastic behavior of fibers is coupled to the failure 
of fibers \cite{kun4}.
For the behavior of the solutions $\varepsilon(t)$ of Eq.\
(\ref{eq:eom}) two distinct 
regimes can be distinguished depending on the value of the external
load $\sigma_{\rm o}$: When $\sigma_{\rm o}$ falls below a critical value
$\sigma_{\rm c}$ Eq.\ (\ref{eq:eom}) has a stationary solution
$\varepsilon_s$, which can be obtained by setting
$\dot{\varepsilon}=0$, {\it i.e.}
$\sigma_{\rm o} = E\varepsilon_s[1-P(\varepsilon_s)]$.
It means that until this equation can be solved for $\varepsilon_s$ 
at a given external load $\sigma_{\rm o}$, the solution  $\varepsilon(t)$ of
Eq.\ (\ref{eq:eom}) converges to 
$\varepsilon_s$  when $t\to \infty$, and the system suffers only
a partial failure. However, when 
$\sigma_{\rm o}$  exceeds the critical value $\sigma_{c}$ no stationary
solution exists, furthermore, $\dot{\varepsilon}$ remains always
positive, which implies that for $\sigma > \sigma_{c}$ the strain of the
system   $\varepsilon(t)$ monotonically increases until the 
system fails globally at a finite time $t_f$. The behavior of
$\varepsilon(t)$ is illustrated in Fig.\ \ref{fig:eps_t} for two
values of $\sigma_o$ below and above $\sigma_c$.
It follows 
from the above argument that the critical value of the load
$\sigma_{c}$ is the static fracture strength of the bundle \cite{sornette3}. 

The creep rupture of the viscoelastic bundle can be interpreted so
that for $\sigma_{\rm 
  o} \leq \sigma_{\rm c}$ the bundle is partially damaged
implying an infinite lifetime $t_f = \infty$ and the emergence of a
stationary macroscopic state,
while above the critical load
$\sigma_{\rm o} > \sigma_{\rm c}$ global failure occurs at a finite
time $t_f$, but in the vicinity of $\sigma_c$ the global failure is
preceded by a long lived stationary state.
The nature of the transition occurring at $\sigma_c$ 
can be characterized by analyzing how the creeping system behaves when
approaching the critical load both from below and above.

For $\sigma_{\rm 
  o} \leq \sigma_{\rm c}$ the fiber
bundle relaxes to the stationary deformation $\varepsilon_s$ through
a gradually decreasing breaking activity. 
It can be shown analytically that $\varepsilon(t)$ has
an exponential relaxation to $\varepsilon_s$ with a characteristic
time scale $\tau$ that depends on the external load $\sigma_0$ as
\begin{eqnarray}
  \label{eq:tau_crit}
  \tau \sim \left(\sigma_{c} - \sigma_{\rm o}\right)^{-1/2}, \label{eq1}\ \ \ 
  \mbox{for} \ \ \ \sigma_{o} < \sigma_{\rm c},
\end{eqnarray}
{\em i.e.}, when approaching the critical point from below the
characteristic time of the
relaxation to the stationary state diverges according to a universal
power law with an exponent $-1/2$ independent of the form of disorder
distribution $P$ \cite{note1}. Above the critical point the lifetime
$t_f$ defines the characteristic time scale of the system which can be
cast in the form \cite{kun4}
\begin{eqnarray}
  \label{critical}
  t_f \sim (\sigma_{\rm o} - \sigma_{\rm c})^{-1/2}\label{eq2}, \qquad \mbox{for} \qquad
  \sigma_{\rm o} > \sigma_{\rm c},
\end{eqnarray}
so that $t_f$ also has a power law divergence at $\sigma_{\rm c}$ with a
universal exponent $-\frac{1}{2}$ like $\tau$ below the
critical point. Hence, for global load sharing $\gamma=0$, the system
exhibits scaling behavior on both sides of the critical point
indicating a continuous transition at the critical load $\sigma_c$. 

In a creeping system, due to the steady external driving, local
overloads build up slowly on the microscopic level when the breaking
threshold of fibers is exceeded by the local deformation. The sudden
breaking of fibers occurring on a time scale much shorter than the
scale of the driving provides the relaxation mechanism which resolves
the overloads in the system. This mechanism consists of sequential
fiber breakings that form an avalanche which either stops (below
the critical point) or continues until the whole system gets destroyed
(above the critical point). 

\begin{figure}
\begin{center}
\epsfig{bbllx=140, bblly=345,bburx=470,bbury=640,file=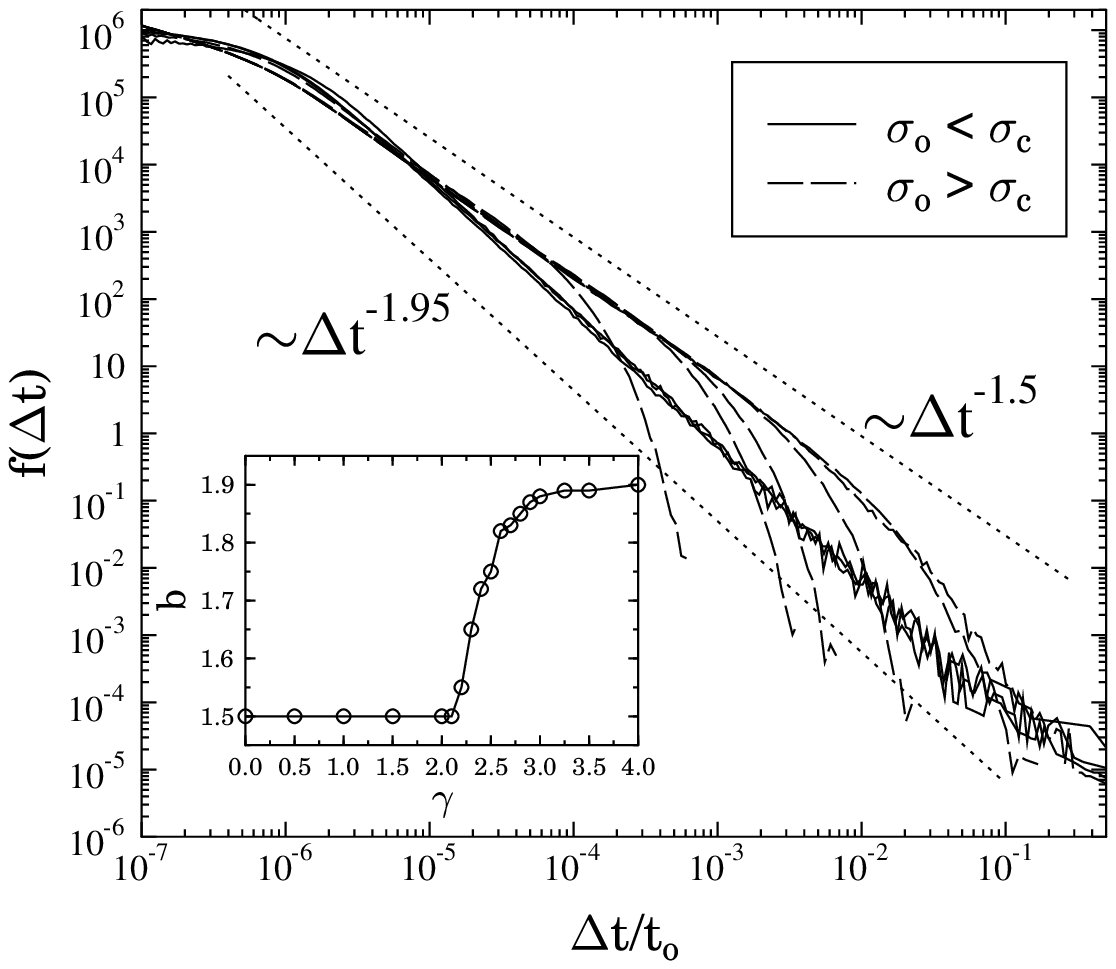,
  width=8cm}  
\caption{\small The distribution of inter-event times $\Delta t$ for
  $\gamma=0$. Power law behavior can be observed over 5 orders of
  magnitude. Inset: the exponent $b$ of $f(\Delta t)$ as a function of
$\gamma$ for $\sigma_o > \sigma_c$.}
\label{fig:delta_dist}
\end{center}
\end{figure}

Fibers fail one-by-one, furthermore, under GLS conditions, breakings
occur in the order of increasing breaking thresholds
$\varepsilon_{i}$ and the time $\Delta t(\varepsilon_i,\varepsilon_{i+1})$
elapsed between the breaking of $i$-th and $i+1$-th fibers can be
analytically obtained. 
The inter-event time $\Delta t$ is a fluctuating quantity which depends
both on the breaking thresholds and the load level, as 
illustrated in the inset of Fig.\ \ref{fig:eps_t}. It can be observed
in the figure 
that before and after the plateau of $\varepsilon(t)$ the inter-event
times are relatively short (low peaks), while along the plateau
$\Delta t$ is scattered over a broad interval.  The statistics of
inter-event times characterized by the distribution $f(\Delta t)$
provides information on the microscopic dynamics of creep. $f(\Delta
t)$ is presented in Fig.\ \ref{fig:delta_dist} for a system of
$N=2\times 10^7$ fibers. Simulations revealed that $f(\Delta t)$
exhibits a power law of the form $f(\Delta t) \sim \Delta t^{-b}$
both below and above the critical point whenever the macroscopic
stationary state 
characterized by the plateau of $\varepsilon(t)$ is attained. However,
the value of the exponent $b$ is different on the two sides of the
critical point, {\it i.e.} below $\sigma_c$ the exponent of the
distribution is $b=1.95\pm 0.05$ independent of $\sigma_o$, while
above $\sigma_c$ we obtained $b=1.5\pm 0.05$. Increasing the load
above $\sigma_c$ the stationary state gradually disappears implying
that the power law regime of $f(\Delta t)$ preceding the exponential
cut-off is getting shorter but the exponent remains the same (see
Fig.\ \ref{fig:delta_dist}).  It follows that the creeping bundle
self-organizes into a critical state: the threshold dynamics of
the system characterized by a separation of time scales of the
external driving and the relaxation process leads to the emergence of
a macroscopic stationary state accompanied by power law distributed
microscopic events.

\begin{figure}
\begin{center}
\epsfig{bbllx=125, bblly=390,bburx=480,bbury=670,file=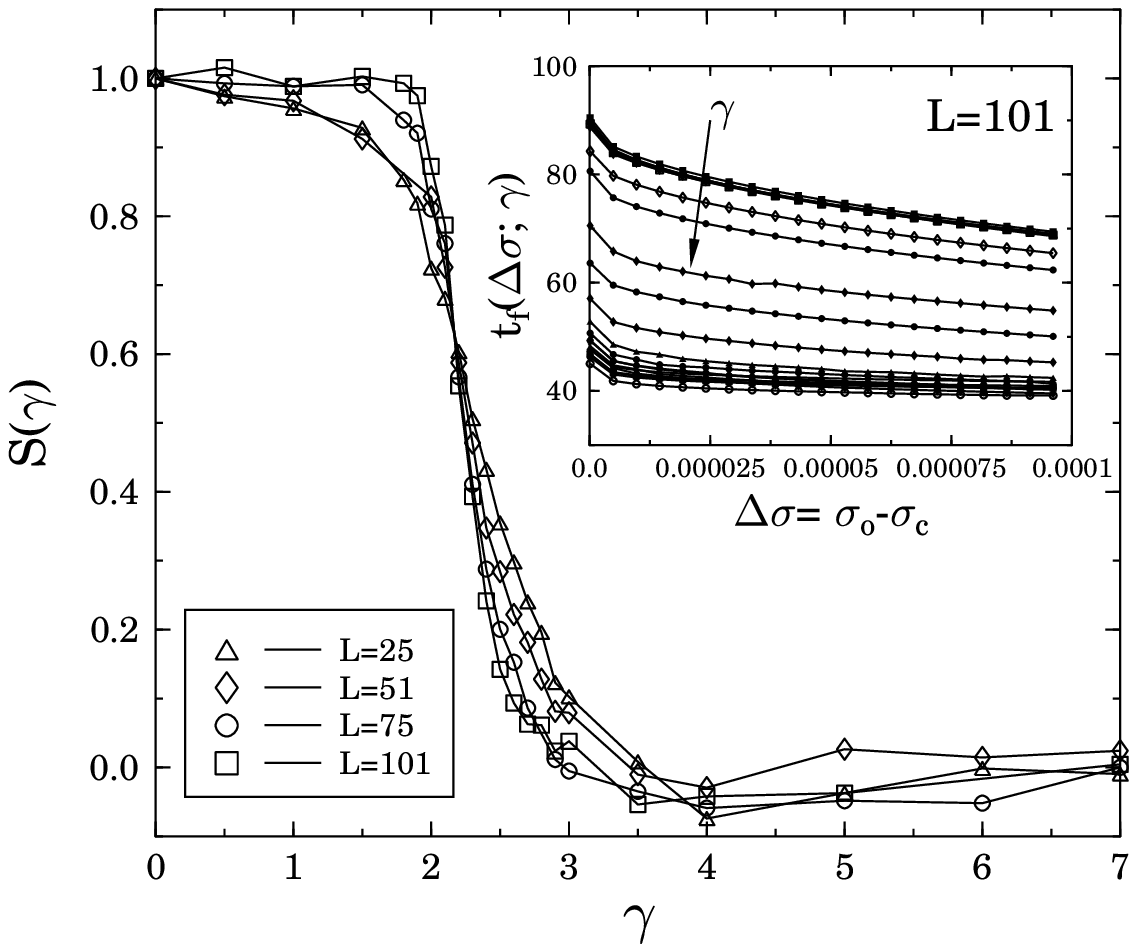,
  width=8.5cm}  
\caption{\small Inset: the lifetime $t_f$ as a function of
  $\sigma_o-\sigma_c$ for different values of the exponent $\gamma$
  between 0 and 100. $S(\gamma)$ is presented in the main figure for
  several system sizes $L$.}  
\label{fig:local_tf}
\end{center}
\end{figure}

To explore the effect of the details of load redistribution on
the creep rupture process we studied how the behavior of the system
changes in the vicinity of the critical point when the load sharing gets
localized. Simulations have  
been performed varying the effective range of interaction of fibers by
controlling the exponent $\gamma$ of the load sharing function. 
The inset of Fig.\ \ref{fig:local_tf} presents the lifetime $t_f$ of a
bundle of fibers arranged on a square lattice of side length $L=101$
as a function of the distance from the critical point $\Delta \sigma =
\sigma_o-\sigma_c$ for several values of the exponent $\gamma$.  It
can be observed that the $t_f(\Delta \sigma;\gamma)$ curves form two
groups of different functional form: The upper group is obtained for
$0 \leq \gamma \leq 1.95$ when the load sharing is global
\cite{yamir2} and Eq.\ (\ref{eq2}) holds \cite{kun4}. However, in the
lower group, obtained for $\gamma > 2.9$ when the load sharing gets
localized \cite{yamir2}, $t_f(\Delta \sigma;\gamma)$ rapidly takes a
constant value showing an abrupt transition at the critical load
$\sigma_c$ with no scaling,
reminiscent of a first order transition. The
results imply the existence of two universality classes in creep
rupture characterized by a completely global (GLS), or a completely
local (LLS) behavior depending on the effective range of interaction
$\gamma$ with a rather sharp transition between them. In order to
quantify the behavior of $t_f(\Delta \sigma;\gamma)$ under the
variation of $\gamma$ we calculated the normalized quantity $S(\gamma)
= \left[ t_f(\gamma)-t_f(\gamma=10)\right]/ \left[
t_f(\gamma=0)-t_f(\gamma=10) \right]$, where $t_f(\gamma)$ denotes the
value of $t_f$ at the smallest value of $\Delta \sigma$ used to
calculate $t_f(\Delta \sigma;\gamma)$ at a given $\gamma$.  Fig.\
\ref{fig:local_tf} shows that $S(\gamma)$ provides a
quantitative description of the creep rupture transition in terms of
the effective range of interaction so that $S(\gamma)$ takes value unity
for the GLS, and it has a value close to zero for the LLS class,
respectively. It can also be observed in Fig.\ \ref{fig:local_tf} that
the transition between the two universality classes gets sharper
around $\gamma_c \approx 2$ with increasing system size.
Real materials
described by a finite value of $\gamma$ must fall into one of the
above universality classes. The existence of only two universality
classes implies that the mean field analytical results can be
extended beyond $\gamma=0$, {\em i.e.}, they apply for a wider
interaction range which is relevant for real materials.

Extensive simulations revealed that whenever a macroscopic stationary
state is attained 
by the system, the distribution of inter-event times follows a power
law irrespective of the range of interaction $\gamma$. Below the
critical point the exponent $b$ of the distribution has 
a value $b=1.95 \pm 0.05$ independent of $\gamma$, while above the
critical point $b$ is different in the two universality classes as 
illustrated by the inset of Fig.\ \ref{fig:delta_dist}. In the LLS
class the 
exponent $b$ has practically the same value below and above
$\sigma_c$.  

The breaking process of fibers occurring in a solid under various
loading conditions can be monitored by acoustic emission techniques
which has also been applied to study creep rupture. The statistics of
inter-event times has been studied in various types of materials like
wood, plaster, basalt, and fiber glass. It was found experimentally
that the distribution of inter-event times always exhibits a power law
behavior, however, the values of $b$ was found to depend on the
material falling between 1.2 and 1.9
\cite{exper1,guarino1,guarino2,aless,exper2,hirata} for $\sigma_o >
\sigma_c$. Hence, our theoretical findings are in quite reasonable
agreement with the available experimental results. Moreover, the
different values of $b$ below and above $\sigma_c$ predicted by our
model for long range interactions would correspond to different
Omori's exponents for foreshocks and aftershocks in earthquake
dynamics, which has also been observed recently \cite{sornette}.

In conclusion, we have identified two universality classes of creep
rupture depending on the range of interaction of fibers. The critical
behavior of the microfracturing process can be seen as the result of
the self-organization of the system into a macroscopic stationary
state whose duration depends on the external perturbation (load). In
this sense the system is at some point of marginal stability jumping
form one metastable state to another with power-law distributed
waiting times. This suggests the existence of a critical dynamics
underlying the process that seems to indicate self-organized
criticality \cite{exper1,aless,bak} in creep. Our theoretical results
provide a consistent explanation of recent experimental findings on
the damage process of creep rupture.

Y.\ M.\ thanks A.\ Vespignani for enlighting discussions and A.\ F.\
Pacheco for useful comments. This work was supported by the project
SFB381, by the NATO grant PST.CLG.977311.  F.\ K acknowledges the kind
hospitality of the Condensed Matter Group of ICTP Trieste, and
financial support of the B\'olyai J\'anos Fellowship of the Hungarian
Academy of Sciences and of the Research Contracts FKFP 0118/2001 and
T037212. Y.\ M.\ acknowledges financial support from the Ministerio de
Educaci\'on Cultura y Deportes (Spain) and of the Spanish DGICYT
Project BFM2002-01798.

\end{multicols}


\begin{thebibliography}{100}

\bibitem{zap98} S. Zapperi, P. Cizeau, G. Durin, and H. E. Stanley,
Phys.\ Rev.\ B {\bf 58}, 6353 (1998).

\bibitem{cha} {\sl Statistical Physics of Fracture and Breakdown in
Disordered Systems\/}. B. K. Chakrabarti, L. G. Benguigui, Clarendon
Press, Oxford (1997), and references therein.

\bibitem{tur97} D. L. Turcotte,{\sl Fractals and Chaos in Geology and
Geophysics}. (2nd Ed. Cambridge University Press, Cambridge, New York,
1997).

\bibitem{exper1} C.\ Maes, A.\ Van Moffaert, H.\ Frederix, and H.\
Strauven, Phys.\ Rev.\ B {\bf 57}, 4987 (1998).

\bibitem{guarino1} A.\ Guarino, A.\ Garcimart\'\i n, and S.\
  Ciliberto, Eur.\ Phys.\ J.\ B {\bf 6}, 13 (1998).

\bibitem{guarino2} A.\ Guarino, S.\ Ciliberto, A.\ Garcimart\'\i n,
  M.\ Zei, and R.\ Scorretti, Eur.\ Phys.\ J.\ B {\bf 26}, 141 (2002).

\bibitem{aless}A.\ Petri, G.\ Paparo, A.\ Vespignani, A.\ Alippi, and
M.\ Constantini, Phys.\ Rev.\ Lett.\ {\bf 73}, 3423 (1994).

\bibitem{exper2} A.\ Vespignani, A.\ Petri, A.\ Alippi, and G.\ Paparo,
Fractals {\bf 3}, 839 (1995).

\bibitem{hirata} T.\ Hirata, J.\ Geophys.\ Res.\  {\bf 92}, 6215 (1987).

\bibitem{hansen} M.\ Kloster, A.\ Hansen, and P.\ C.\ Hemmer, Phys.\
  Rev.\ E {\bf 56}, 2615 (1997).

\bibitem{sornette3}J.\ V.\ Andersen, D.\ Sornette, and K.-T.\ Leung,
  Phys.\ Rev.\ Lett.\ {\bf 78}, 2140 (1997).
\bibitem{kun1} F.\ Kun, S.\ Zapperi, and H.\ J.\ Herrmann, 
       European Physical Journal B{\bf 17}, 269 (2000).
\bibitem{kun4} R.\ C.\ Hidalgo, F.\ Kun, and H.\ J.\ Herrmann,
  Phys.\ Rev.\ E {\bf 65}, 032502 (2002).
\bibitem{yamir1} Y.\ Moreno, J.\ B.\ G\'{o}mez, A.\ F.\ Pacheco, 
       Phys.\ Rev.\ Lett.\ {\bf 85}, 2865 (2000). 

\bibitem{yamir2} R.\ C.\ Hidalgo, Y.\ Moreno, F.\ Kun, and H.\ J.\
  Herrmann, Phys.\ Rev.\ E {\bf 65}, 046148 (2002). 

\bibitem{chakrab} S.\ Pradhan and B.\ K.\ Chakrabarti, Phys.\
Rev.\ E {\bf 65}, 016113 (2001); S.\ Pradhan, P.\ Bhattacharyya, and
B.\ K.\ Chakrabarti, Phys.\ Rev.\ E {\bf 66}, 016116 (2002).

\bibitem{moral} L. Moral, Y. Moreno, J. B. G\'{o}mez, A. F. Pacheco,
Phys. Rev. E {\bf 63}, 066106 (2001).

\bibitem{newzar} W. I. Newman, A. M. Gabrielov, T. A. Durand,
S. L. Phoenix, and D. L. Turcotte, Physica D {\bf 77}, 200 (1994);
Y. Moreno, A. M. Correig, J. B. G\'{o}mez, A. F. Pacheco,
J. Geophys. Res. {\bf B 106}, 6609 (2001).

\bibitem{omori} F.\ Omori, Rep.\ Eart. Inv. Comm. {\bf 2}, 103 (1894);
  T.\ Utsu, Fac.\ Science, Hukkaido University {\bf 3}, 129 (1969).

\bibitem{note1} Note that a similar power law divergence of the number
of successive relaxation steps of a dry fiber bundle subjected to a
constant external load was pointed out in Ref.\ \cite{chakrab}.

\bibitem{sornette} A. Helmstetter, D. Sornette, and J.-R. Grasso,
preprint cond-mat/0205499 (2002).

\bibitem{bak} P.\ Bak, C.\ Tang, and K.\ Wiesenfeld, Phys.\ Rev.\ Lett.\
{\bf 59}, 381 (1987).

\end{thebibliography}
\end{document}